\documentclass[groupedaddress]{revtex4}

\usepackage{graphicx}
\usepackage{natbib}
\usepackage{amssymb}
\usepackage{amsmath}

\begin{document}

\title{A Note on the Dipole Coordinates}
\author{Akira Kageyama}
\email{kage@jamstec.go.jp}
\author{Tooru Sugiyama}
\author{Kunihiko Watanabe}
\author{Tetsuya Sato}
\affiliation{Earth Simulator Center,
Japan Agency for Marine-Earth Science and Technology,
Yokohama 236-0001, Japan}

\begin{abstract}
A couple of orthogonal coordinates for dipole geometry
are proposed for numerical simulations of plasma geophysics
in the Earth's dipole magnetic field.
These coordinates have proper metric profiles
along field lines 
in contrast to the standard dipole coordinate system
that is commonly used in analytical studies for dipole geometry.
\end{abstract}

\maketitle

%**************************************************
\section{Introduction}
%**************************************************
In the study of plasma geophysics,
an orthogonal coordinate system defined by a dipole field is
commonly used because of the Earth's dipole magnetic field $\mathbf{B}_d$.
The standard dipole coordinate system $(\mu,\chi,\phi)$ 
is defined through the spherical coordinates
$(r,\theta,\phi)$ as
\begin{equation}				\label{eq:000a}
 \mu  =   -\frac{\cos{\theta}}{r^2}, 
\qquad
 \chi =    \frac{\sin^2{\theta}}{r},
\end{equation}
where $r$ is length from Earth's center,
normalized by its radius $1Re$,
$\theta$ is colatitude,
and $\phi$ is the longitude.
The coordinate $\mu$ is a potential function of a dipole field,
$\mathbf{B}_d\propto \nabla \mu$,
and constant--$\chi$ curves in a meridian 
plane, $\phi=\mbox{const.}$, denote dipole filed lines.

Since $(\mu,\chi,\phi)$ is an orthogonal system,
their metric terms are simply given by
\begin{align}				
 h_\mu &= 1 \,/\, |\nabla \mu| = r^3\,/\,\Theta,	\label{eq:1001}\\
 h_\chi &= 1 \,/\, |\nabla \chi| 
		= r^2\,/\,(\Theta\, \sin\theta),	\label{eq:1002}\\
 h_\phi &= 1 \,/\, |\nabla \phi| = r\,\sin\theta,	\label{eq:1003}
\end{align}
with
\begin{equation}
   \Theta(\theta) = \sqrt{1+3\,\cos^2\theta},
\end{equation}
and the length element $ds$ is given by
$ds^2=ds_\mu^2+ds_\chi^2+ds_\phi^2$ with
$ds_\mu=h_\mu\,d\mu$,
$ds_\chi=h_\chi\,d\chi$,
$ds_\phi=h_\phi\,d\phi$.
Given these metric terms,
it is straightforward to discretize
any differential operator
such as the divergence of a vector
$\mathbf{v}$ that is denoted by components $\{v_\mu,v_\chi,v_\phi\}$ 
in the dipole coordinates as
\begin{equation}					\label{eq:437}
\nabla \cdot \mathbf{v} = 
      \frac{1}{h_\mu h_\chi h_\phi}\frac{\partial}{\partial \mu}
                  (h_\chi h_\phi v_\mu)
    + \frac{1}{h_\mu h_\chi h_\phi}\frac{\partial}{\partial \chi}
                  (h_\phi h_\mu v_\chi)
    + \frac{1}{h_\mu h_\chi h_\phi}\frac{\partial}{\partial \phi} 
		  (h_\mu h_\chi v_\phi),
\end{equation}
by, for example, 
a finite difference method in the computational $(\mu,\chi,\phi)$ space.

The above standard dipole coordinates is convenient 
and certainly appropriate 
for analytical studies in which the Earth's dipolar field plays
central roles.
It also works as a base coordinates for the node and cell
generation of the finite element method in the
dipole geometry \citep{fujita_2000,fujita_2002}.
However, when one tries to use other numerical methods
in which analytical expression of the metric terms
are important for preserving numerical simplicity and accuracy,
as in the case of the finite difference method,
the standard dipole coordinate $(\mu,\chi,\phi)$ cannot
be used in its original form since
the metric $h_\mu$ changes intensely along the field lines.

It should be noted that $h_\mu\propto |\mathbf{B}_d|^{-1}$
from the above definitions,
which means that
$h_\mu$ is roughly proportional to $r^3$.
Therefore, the metric $h_\mu$ at $r=1$
is $O(10^3)$ smaller than that at $r=10$.
Fig.\,\ref{fig:metric_distrib}(a) shows the $h_\mu$ profile
along a field line starting from $70^\circ$N
as a function of $\mu$.
(We suppose that the north pole is located in $\theta=0$ in this note.)
This field line goes through the equator ($\mu=0$) at $r=8.55$.
Note the sharp peak in Fig.\,\ref{fig:metric_distrib}(a)
at the equator.

\begin{figure}
\begin{center}
\includegraphics[width=0.45\linewidth]{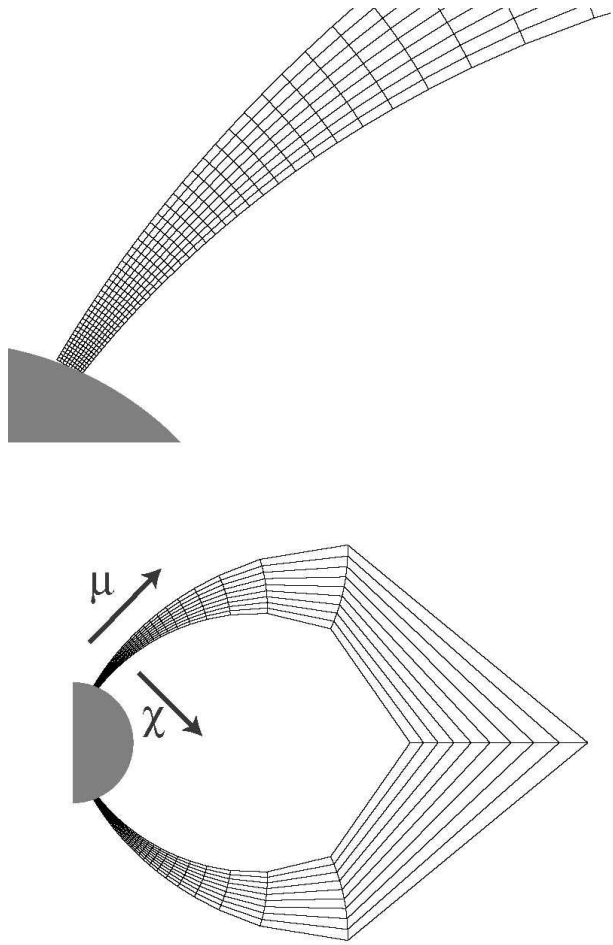}
\end{center}
\caption{\label{fig:grid_standard}
The standard dipole coordinates $(\mu,\chi,\phi)
=(-\cos\theta\,/\,r^2,\sin^2\,/\,r,\phi)$
in a meridian plane, $\phi=\mbox{const.}$
The grid points are distributed with equal spacings
in each direction in the computational space $\mu$ and $\chi$.
The total grid size is $N_\mu\times N_\chi=101\times 10$.
There is no skip of grid points in the figure.
The starting points of the field lines are
between $65^\circ$N and $70^\circ$N at $r=1$.
The upper panel is a closer view.
}
\end{figure}

When one uses the finite central difference method,
the grid spacing along the field line
is given by $\Delta s_\mu=h_\mu\,\Delta \mu$.
Fig.\,\ref{fig:grid_standard} shows
grid point distribution
in the standard dipole coordinates.
The grid size in the figure is $N_\mu\times N_\chi=101\times 10$.
($101$ grids along each field line and
$10$ grids in the perpendicular direction.)
The starting points of the field lines are between
$65^\circ$N and $70^\circ$N at $r=1$.
All the grid points are shown in the figure without any skip.
It is clearly seen that the 
resolution near the equator is so poor that
any numerical simulation on this grid system is impractical.
% ---<added in ver.2>---
For the field line starting from $70^\circ$N at $r=1$,
the metric $h_\mu$ on the equator 
is about $1160$ times larger than that on $r=1$.
% ---<added in ver.2>---
Also note that the imbalance of the grid spacings
between the near Earth and the near equatorial regions
along the field lines causes unnecessarily severe
restriction on the 
Courant-Friedrichs-Lewy condition in explicit time integration schemes.

For some numerical simulations such as the
magnetosphere-ionosphere coupling,
it is certainly desirable to use a grid system
that has a natural grid convergence
near the ionosphere to resolve fine structures near the coupling region,
but the three-orders of magnitude is obviously too much.
This is especially serious
when one tries to simulate some phenomena in which
relatively high resolution near the equator
is required.
An example of such simulation is the
auroral arc formation by the feedback instability
driven by vortex flow in the equator \citep{watanabe_1988,watanabe_1993}.

A trivial way to avoid the poor resolution problem
of the standard dipole coordinates near the equator
is to place the computational grid points
along the $\mu$ space in a nonuniform way.
In this case, 
the metric factors have to be numerically calculated.
For example, \citet{lee_1989,lee_1991} determined
the grid spacing due to the local Alfv\'en wave speed.
The same approach was adopted in \citet{budnik_1998}.
However, this method injures the generality
of the dipole coordinates as well as its 
analytical nature.

If one prefers to fully numerical methods,
refer to \citet{proehl_2002} in which
a general algorithm
to construct grid points along an arbitrarily given
magnetic field, including the dipole, is presented.
In contrast to that approach,
we propose in this note
analytical as well as simple coordinate transformations
of $\mu$ that lead to practical metric distributions
along the field line.

%**************************************************
\section{Transformation formula of the dipole coordinates}
%**************************************************
Before we go into the description on the
modified dipole coordinates defined by the coordinate transformation
of the standard dipole coordinates,
we derive analytical expressions
of the inverse transformation from
the standard dipole coordinates $(\mu,\chi,\phi)$
into the spherical coordinates $(r,\theta,\phi)$
since we could not find these expressions in the literature
and they can be directly applied to
the modified dipole coordinates described later.

Eliminating $r$ from in eq.~(\ref{eq:000a})
with subsidiary variables $u$ and $\zeta$ defined as
\begin{equation}
  u=\sin^2\theta,
  \quad
  \zeta = (\mu/\chi^2)^2,
\end{equation}
we get a fourth order equation of $u$:
\begin{equation}				\label{eq:8000}
\zeta\,u^4+u-1=0.
\end{equation}
The unique solution of eq.~(\ref{eq:8000}) for positive real $u$ is
\begin{equation}				\label{eq:007}
  u = -\frac{1}{2}\,\sqrt{w} +
       \frac{1}{2}\,\sqrt{-w+\frac{2}{\zeta\sqrt{w}}},
\end{equation}
where
\begin{equation}				\label{eq:008}
   w(\zeta) = -\frac{c_1}{\gamma(\zeta)} 
             + \frac{\gamma(\zeta)}{c_2\,\zeta},
\end{equation}
\begin{equation}				\label{eq:010}
  c_1 = 2^{7/3}\,3^{-1/3},
  \quad
  c_2 = 2^{1/3}\,3^{2/3},
\end{equation}
and
\begin{equation}				\label{eq:009}
   \gamma(\zeta) = \left(
		9\,\zeta + \sqrt{3}\, \sqrt{27\,\zeta^2+256\,\zeta^3}
            \right)^{1/3}.
\end{equation}
The analytical expression for $r$ and $\theta$ by
$\mu$ and $\chi$ are, therefore, given by the function $u$:
\begin{equation}				\label{eq:013}
      r(\mu,\chi) = u\, /\, \chi,
\end{equation}
\begin{equation}				\label{eq:012}
      \theta(\mu,\chi) = \arcsin{\sqrt{u}},
\end{equation}
where $\arcsin$ is defined as a continuous function of $u$
with the range of $[0,\pi]$.

%**************************************************
\section{Modified Dipole Coordinates}
\begin{figure}
\begin{center}
\includegraphics[width=0.45\linewidth]{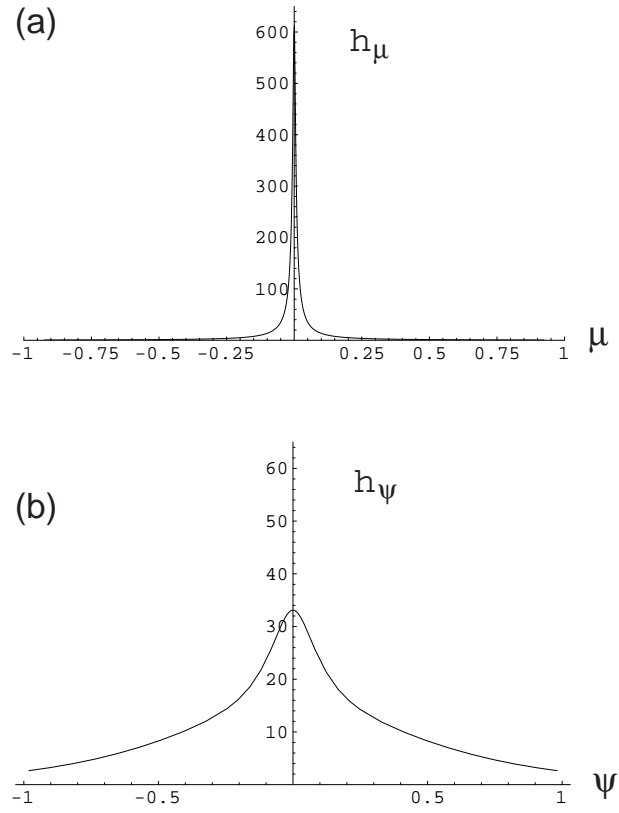}
\end{center}
\caption{\label{fig:metric_distrib}
The metric distribution along field the line starting
from $70^\circ$N at $r=1$.
(a) $h_\mu$ for the standard dipole coordinates
(b) $h_\psi$ for the modified dipole coordinates.
Note that the vertical scales between the two
panels (a) and (b) are different for one order of magnitude.
}
\end{figure}

%**************************************************
The problem of the metric imbalance along field lines
in the standard dipole coordinates originates
from the power $2$ of the
$\mu$'s denominator $r^2$ in eq.~(\ref{eq:000a}).
Therefore, one simple idea
to reduce the steep metric distribution 
in the standard dipole coordinates 
shown in Fig.\,\ref{fig:metric_distrib}(a)
is to use a coordinates $(\mu',\chi,\phi)$ in which
field-aligned coordinate $\mu'$, instead of $\mu$,
is defined as
\begin{equation}				\label{eq:700}
 \mu'=-\frac{\sqrt{\cos\theta}}{r}, \qquad \hbox{for}\quad \theta < \pi/2.
\end{equation}
It is easy to confirm that
 $(\mu',\chi,\phi)$ is also an orthogonal system.
The metric of $\mu'$-coordinate is given by
\begin{equation}				\label{eq:701}
 h_{\mu'} = 1 \,/\, |\nabla \mu'| 
                  = 2\,r^2\,\sqrt{\cos\theta} \,/\, \Theta.
\end{equation}
\citet{hysell_2002} used essentially the same
coordinates as $(\mu',\chi,\phi)$ for a plasma clouds simulation
in midlatitude.
(They used $M\equiv r/\sqrt{\cos\theta}=-1/\mu'$, instead of $\mu'$.)
%
% ----<old>----
%A drawback of this coordinates $(\mu',\chi,\phi)$ 
%in general dipole geometry is
%that the metric $h_{\mu'}$ vanishes in the equator;
%$\left. h_{\mu'}\right|_{\mu'=0}=0$.
%Therefore, the coordinate system $(\mu',\chi,\phi)$ cannot
%be applied for the full range of latitude.
% ----</old>----
%
% ----<new>----
When one uses this coordinate system $(\mu',\chi,\phi)$,
a care should be taken for the fact that
the metric $h_{\mu'}$ vanishes in the equator;
$\left. h_{\mu'}\right|_{\mu'=0}=0$.
% ----</new>----
%
This was not a problem
in the simulation by \citet{hysell_2002} since
it was sufficient for them to use only a small part of the $\mu'$ space
($-0.79\le\mu'\le -0.74$).

\begin{figure}
\begin{center}
\includegraphics[width=0.45\linewidth]{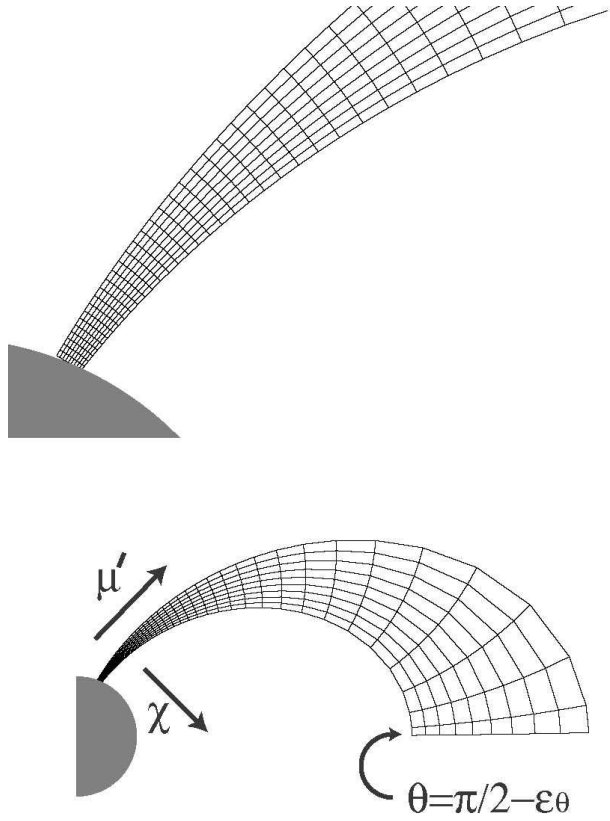}
\end{center}
\caption{\label{fig:modified_dipole_sqrt}
Modified orthogonal dipole coordinates $(\mu',\chi,\phi)$
in a meridian plane.
The coordinate $\mu'$ is defined as $\mu'=-\sqrt{\cos\theta}\,/\,r$,
with $\mu'\le 0-\epsilon_{\mu'}$,
or $\theta \le \pi/2-\epsilon_\theta$.
The small buffer $\epsilon_{\mu'}$ or $\epsilon_\theta$ is
introduced to avoid the vanishing metric in the equator.
($h_{\mu'}=0$ at $\mu'=0$.)
Here $\epsilon_\theta=0.01$. 
The total grid size is $N_{\mu'}\times N_\chi=51\times 10$
in this ``almost'' northern hemispheric region.
}
\end{figure}

%
% ----<old>----
%In spite of the singularity of $h_{\mu'}$ in the equator,
%we point out that one can extend
%the $\mu'$-space to, at least, the 
%full range of a hemisphere (e.g., the northern hemisphere)
%by setting an upper limit of $\mu'$ in such a way as
%$\mu' \le 0-\epsilon_{\mu'}$,
%or $\theta\le\pi/2-\epsilon_\theta$
%with a small positive buffer $\epsilon_{\mu'}$ or $\epsilon_\theta$.
% ----</old>----
%
% ----<new>----
A simple remedy to avoid the singularity of $h_{\mu'}$ in the equator
is not to place the grid point just on the equator.
For example, 
the northern hemisphere is fully covered in practice by 
$\mu' \le 0-\epsilon_{\mu'}$,
or $\theta\le\pi/2-\epsilon_\theta$,
with a small positive buffer $\epsilon_{\mu'}$ or $\epsilon_\theta$.
% ----</new>----
%
Fig.\,\ref{fig:modified_dipole_sqrt} shows the grid
points with $N_{\mu'}\times N_\chi=51\times 10$
for the practically full range of the northern hemisphere
by setting $\epsilon_\theta=0.01$.

The transformation formula for
the modified dipole coordinates $(\mu',\chi,\phi)$
into the spherical coordinates
are obtained by the same equation~(\ref{eq:8000})
by letting
$\zeta = (\mu' / \chi)^4$.

Another form of modified orthogonal dipole coordinates 
proposed in this note is $(\psi,\chi,\phi)$, where
the new coordinate $\psi$ is defined through $\mu$ as
\begin{equation}				\label{eq:800}
 \psi = \sinh^{-1}{(a\,\mu)}\,/\,\bar{a},
\end{equation}
or its inverse transformation
\begin{equation}				\label{eq:801}
 \mu = \sinh{(\bar{a}\,\psi)}\,/\,a,
\end{equation}
where $a$ is a parameter that controls the
grid distribution along dipole field lines,
and $\bar{a}$ is defined as $\bar{a}=\sinh^{-1}{a}$.
Note the identity of $\sinh^{-1}{x}=\log{\left(x+\sqrt{1+x^2}\right)}$.

The metric of $\psi$ is given by
\begin{equation}
 \label{eq:1947}
  h_\psi = h_\mu\,\frac{d\mu}{d\psi}
         = \bar{a}\,r^3\,\cosh{(\bar{a}\,\psi)} \,/\, (a\,\Theta).
\end{equation}

The $h_\psi$ distribution as a function of $\psi$
when the control parameter $a=100$ 
is shown in Fig.\,\ref{fig:metric_distrib}(b),
which should be compared with
$h_\mu$ distribution shown in Fig.\,\ref{fig:metric_distrib}(a).
It should be noted that the vertical scales
in Fig.\,\ref{fig:metric_distrib}(a) and (b) are
different for one order of magnitude.
The basic idea that has lead to the transformation~(\ref{eq:800})
is to relax the steep gradient of the metric along $\mu$
in Fig.\,\ref{fig:metric_distrib}(a)
by local scale transformations.
% --<added in ver.2>---
We want to make the $\mu$ grid spacing along the field 
line being ``shrunk'' only near 
the equator; $\mu\sim0$ or $\psi\sim0$.
Therefore, the grid ``shrink rate'' along the field line
should be a function
with a steeple-like peak at $\psi=0$.
An example of such a function is $1/\cosh \psi$.
The grid shrink rate is given by $d\psi/d\mu$ since it is the reciprocal 
of the grid ``stretch rate'' $d\mu/d\psi$; see eq.~(\ref{eq:1947}).
Solving $d\psi/d\mu = 1/\cosh \psi$, we get
$\mu=\sinh\psi$.
% --</added in ver.2>---

In the limit of $a\rightarrow 0$,
$\psi=\mu$.
As the parameter $a$ increases, 
grid points near the Earth ($r=1$) 
along field lines,
which are highly concentrated 
in the standard dipole coordinates
(see the upper panel of Fig.\,\ref{fig:grid_standard}),
move toward the equator along the field lines.
The denominator $\bar{a}$ in eq.~(\ref{eq:800}) is introduced
to keep the transformed coordinate $\psi$
being always in the range of $[-1,1]$.
Fig.\,\ref{fig:modified_dipole} shows grid points
distribution when $a=100$ 
for $N_\psi\times N_\chi=101\times 10$.
% ---<added in ver.2>---
The metric $h_\psi$ for the field line starting from $70^\circ$N
on the equator 
is only $12$ times larger than that on $r=1$.
% ---</added in ver.2>---
The coordinate transformation 
by the $\sinh$-function---applied to the cartesian coordinates---was 
also used
in our numerical simulations of 
the magnetosphere \citep{kageyama_1992,usadi_1993}.

The coordinates transformations from $(\psi,\chi,\phi)$ into 
$(r,\theta,\phi)$ are given by eqs.~(\ref{eq:013}) and~(\ref{eq:012})
with eq.~(\ref{eq:801}).

The relation between components of a vector $\mathbf{v}$
in the spherical coordinates $\{v_r,v_\theta,v_\phi\}$
and in the modified dipole coordinates $\{v_\psi,v_\chi,v_\phi\}$ is
given by the same form as that in the standard dipole coordinates:
\begin{equation}
\left(
\begin{array}{c}
v_\psi\\
v_\chi\\
v_\phi
\end{array}
\right)
=
\left(
\begin{array}{ccc}
  2\cos{\theta}\,/\,\Theta
& \sin{\theta}\,/\,\Theta
& 0 \\
  -\sin{\theta}\,/\,\Theta
& 2\cos{\theta}\,/\,\Theta
& 0 \\
  0
& 0   
& 1
\end{array}
\right)
\left(
\begin{array}{c}
v_r\\
v_\theta\\
v_\phi
\end{array}
\right).
\end{equation}
The inverse transformation is given by the transverse matrix.

%**************************************************
\section{Discussion and Summary}
\begin{figure}
\begin{center}
\includegraphics[width=0.45\linewidth]{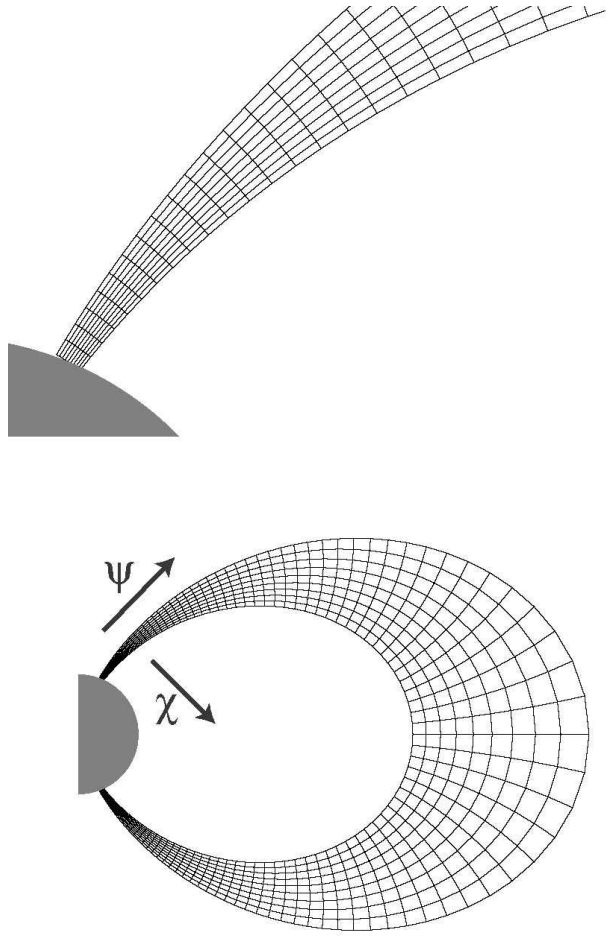}
\end{center}
\caption{\label{fig:modified_dipole}
Modified orthogonal dipole coordinates $(\psi,\chi,\phi)$
in a meridian plane.
The coordinate $\psi$ is defined as
$\psi=\sinh^{-1}{(a\,\mu)}\,/\,\sinh^{-1}{a}$.
Total grid size is $N_\psi\times N_\chi=101\times 10$.
The control parameter $a=100$ in this figure.
Compare with the grid distribution
of the standard dipole coordinates (Fig.\,\ref{fig:grid_standard})
with the same grid size.
}
\end{figure}

%**************************************************
The standard dipole coordinates $(\mu,\chi,\phi)$ defined 
by eq.~(\ref{eq:000a}) is not a
good choice for a base grid in numerical studies since
the metric contrast along each field line is too intense.
Instead of the standard dipole coordinates,
we propose to use the modified orthogonal dipole coordinates
defined by
\begin{equation}			\label{eq:888}
 (\psi,\chi,\phi) =
  \left(
         -\frac{\sinh^{-1}{(a\cos\theta/r^2)}}{\sinh^{-1}{a}},
          \frac{\sin^2\theta}{r},
          \phi
   \right),
\end{equation}
with a tuning parameter $a$ of the
metric distribution along field lines.
%
% --<added in ver.2>--
The standard dipole coordinates
is a special case in the limit of $a\rightarrow 0$.
Fig.\,\ref{fig:modified_dipole} shows the case when $a=100$
for the field lines with foot points located between $65^\circ$N and $70^\circ$N.
Since $\psi$'s metric, or the grid distribution, is not 
sensitive to the change of $a$ around this value,
one would not have to perform its fine control.
For other $\chi$ range, pertinent $a$ value will be
easily found by visual checks of grid distribution images
like Fig.\,\ref{fig:modified_dipole}.
% --</added in ver.2>--
%

For problems in which a symmetry around the equator
between the northern and southern hemispheres is present,
one can also try another form of modified 
orthogonal dipole coordinates defined by
\begin{equation}
 (\mu',\chi,\phi) =
  \left(
        -\frac{\sqrt{\cos\theta}}{r},
         \frac{\sin^2\theta}{r},
         \phi
  \right),
\end{equation}
in which $\mu'\le 0-\epsilon$, with a small positive buffer $\epsilon$
for the northern hemisphere.

Recently, a non-orthogonal dipole coordinates
that is designed so that
the lower-most constant-$\mu$ surface
coincides with a constant-$r$ surface (i.e., a sphere)
is presented \citep{lysak_2004}.
It is straightforward and effective 
to apply the coordinate transformations
presented in this note
for that nonorthogonal dipole coordinates, too.

%**************************************************

%- %\bibliographystyle{elsart-harv}  This doesn't work.... Too many ref?
%- \bibliographystyle{apalike}
%- \bibliography{without_mymemo}

%**************************************************

\newpage

%% ------------------------------------------------------------------------ %%
%
% FIGURES
%
% ---------------

%--------------------------figure-----------------------------------------
%--------------------------figure-----------------------------------------

%--------------------------figure-----------------------------------------
%--------------------------figure-----------------------------------------

%--------------------------figure-----------------------------------------
%--------------------------figure-----------------------------------------

%--------------------------figure-----------------------------------------
%--------------------------figure-----------------------------------------

\end{document}